\documentclass[12pt]{article}
\usepackage{amsmath,amssymb,amsfonts,amsthm,amsbsy}
\usepackage{graphics}

\newtheorem{theorem}{Theorem}

\newtheorem{lemma}[theorem]{Lemma}

\newtheorem{definition}[theorem]{Definition}

\def\R{\mathbb{ R}}

\title{A Note on Sparsification by Frames}

\author{ Christopher A Baker\\
Dept. of EE \& CS, Univ. of Wisconsin-Milwaukee\\
Milwaukee, WI 53211\\
email: {\tt cabaker2@uwm.edu }}

\date{December 18, 2014}
\begin{document}
\maketitle
\begin{abstract}
The purpose of this note is to establish a new generalized Dictionary-Restricted Isometry Property (D-RIP) sparsity bound constant for compressed sensing.
 For fulfilling D-RIP, the constant $\delta_k$ is used in the definition:
$(1 -\delta_k)\|D v\|_2^2 \le \|\Phi D v\|_2^2 \le (1 + \delta_k)\|D  v\|^2$. We prove that signals with $k$-sparse $D$-representation can be reconstructed if
$\delta_{2k} < \frac{2}3$.
 The approach in this note can be extended to obtain other D-RIP bounds (i.e., $\delta_{tk}$).
\end{abstract}


Let $\Phi\in \R^{n\times p}$ and $\beta\in \R^p$ be a signal such that
\[
y=\Phi \beta + z
\]
with $\|z\|_2 \le \varepsilon$. In compressed sensing, one can find a good stable approximation (in terms of $\varepsilon$ and
the tail of $\beta$ consisting of $p-k$ smallest entries) of
$\beta$ from the measurement matrix $\Phi$ and the measurement $y$ through solving an $\ell_1$-minimization, provided that $\Phi$
belongs to a family of well behaved matrices. A subclass of this family of matrices can be characterized by the well known
{\sl restrictive isometry property (RIP)} of Cand\`es, Romberg, and Tao, \cite{CRT,CanTao05}. This property requires  the following relation
for $\Phi$
\[
\sqrt{1-\delta_k}\|c\|_2 \le \|\Phi c\|_2 \le \sqrt{1+\delta_k}\|c\|_2
\]
for every $k$-sparse vector $c$ (namely, $c$ has at most $k$ nonzero components), for some small constant  $\delta_k$. Some
bounds on $\delta$ have been determined, e.g., \cite{CRT,CanTao05,CWX,Candes}. Cai and Zhang recently have established
several sharp RIP bounds
that cover the most interesting cases of $\delta_k$ and $\delta_{2k}$ \cite{CZ,CZ1}, showing $\delta_k\le \frac{1}3, \delta_{2k}<\frac{\sqrt{2}}2$.

The requirement of a signal being sparse or approximately sparse is a key in this setting. Many families of integrating signals
indeed have sparse representations under suitable bases. Recently an interesting sparsifying scheme was proposed by
Candes, Eldar,  Needel, and Randall \cite{CanDRIP}. In their scheme, instead of bases, tight frames are used to sparsify
signals.

Let $D\in \R^{p\times d}$ ($d\ge p$) be a tight frame and $k \le d$. \cite{CanDRIP} suggests that one use the following
optimization to approximate the signal $\beta$:
\begin{equation}\label{eq:1.1}
\hat\beta = \mbox{argmin}_{\gamma\in \R^p} \|D^*\gamma\|_1\quad \mbox{ subject to } \quad \|y - \Phi \gamma \|_2\le \varepsilon.
\end{equation}

The traditional RIP is no longer effective in the generalized setting. Candes, Eldar,  Needel, and Randall defined
the {\sl D-restricted isometry property} which extends RIP \cite{CanDRIP}. Here we shall use the formulation of D-RIP in \cite{LLY}
\begin{definition}
The measurement matrix $\Phi$ obeys the  D-RIP with constant $\delta_k$ if
\begin{equation}\label{eq:1.2}
(1-\delta_k)\|Dv\|_2^2\le \|\Phi D v\|_2^2\le (1+\delta_k)\|Dv\|_2^2
\end{equation}
holds for all $k$-sparse vector $v\in \R^d$.
\end{definition}

\cite{CanDRIP,LLY} have determined some bound for the D-RIP constant $\delta_{2k}$. The purpose of this note is to remark,
using the clever ideas of Cai and Zhang \cite{CZ1}, that one can get an improved bound  for D-RIP constant $\delta_{2k}$
without much difficulty.

\begin{theorem}\label{thm:1.1}
Let $D$ be an arbitrary tight frame and let $\Phi$ be a measurement matrix satisfying
D-RIP with $\delta_{2k} < \frac{2}{3}$. Then the solution $\hat{\beta}$ to (\ref{eq:1.1}) satisfies
\[
\| \beta - \hat{\beta}\|_2 \le C_0\varepsilon + C_1\frac{\|D^*\beta-(D^*\beta)_{\max (k)}\|_1}{\sqrt{k}}
\]
where $C_0, C_1$ are constants that depend on $\delta_{2k}$,  $(D^*\beta)_{\max(k)}$  is the vector
$D^*\beta$ with all but the $k$ largest components (in magnitude) set to zero.
\end{theorem}

Before proving this theorem, let us make some remarks. Firstly, Cai and Zhang have obtained a sharp bound $\delta_{2k}<\frac{\sqrt{2}}2$
 for the case $D=I$ in \cite{CZ1}, it is interesting to know whether the bound of this note can be further improved. Secondly, following the
ideas of \cite{CZ,CZ1}, more general results (other D-RIP bounds) can 
be obtained in parallel.

We need the following $\ell_1$-norm invariant convex $k$-sparse decomposition of Xu and Xu \cite{XuXu}, and Cai and Zhang \cite{CZ1}
 in our proof of theorem \ref{thm:1.1}. We shall take the description from \cite{XuXu}.
\begin{lemma}\label{lem:1.1}
For positive integers $k\le n$, and positive constant $C$,
let $v\in \R^{n}$ be a vector with $\|v\|_1\le C$ and $\|v\|_{\infty}\le \frac{C}k$.
Then there are $k$-sparse vectors $w_1,\dots, w_M$ with
\[
\|w_t\|_1=\|v\|_1 \quad \mbox{ and} \quad \|w_t\|_{\infty}\le \frac{C}k \quad \mbox{ for } t=1,\cdots, M,
\]
such that
\[
v = \sum_{t=1}^M x_t w_t
\]
for some nonnegative real numbers $x_1, \dots, x_M$ with $\sum_{t=1}^Mx_t=1$.
\end{lemma}

Now let us proceed to the proof of theorem \ref{thm:1.1}.
\begin{proof} In this proof we follow the ideas in the proofs of Theorems 1.1 and 2.1 of \cite{CZ1}, incorporating
some more simplified steps. We also
use some strategies from \cite{CWX,CXZ}.
We only deal with the $\delta_{2k}$ case so that the key ideas can be conveyed clearly.

Let $h=\hat{\beta}-\beta$.

For a subset $S\subset \{1,2,\cdots, d\}$, we will denote by $D_S$ the matrix $D$ restricted to the columns indexed by $S$ (and replacing other columns by zero vectors).
Let $\Omega$ denote the index set of the largest $k$ components of $D^*\beta$ (in magnitude), i.e., $(D^*\beta)_{\max(k)}=D_{\Omega}^*\beta$. With this
notation we have $D^*_{\Omega^C}\beta = D^*\beta-(D^*\beta)_{\max (k)}$.
As in \cite{CanDRIP}, one can easily verify
\begin{enumerate}
\item $\|D^*_{\Omega^C}h\|_1\le 2\|D^*_{\Omega^C}\beta\|_1+\|D^*_{\Omega}h\|_1$;
\item $\|\Phi h\|_2 < 2\varepsilon$.
\end{enumerate}

Denote $v_i=\langle D_i, h \rangle$ for $i=1,\cdots,d$, where $D_i$ is the $i$-th column of $D$,  then
\[
D^*h=(v_1,\dots,v_d)^{\top}.
\]
By rearranging the columns of $D$ if necessary, we may assume $|v_1|\ge |v_2|\ge\cdots \ge |v_d|$. Let $T=\{1,2,\dots,k\}$. In this case, we have
\begin{equation}\label{eq:1.3}
D_{T}^*h=(v_1,\dots,v_k, 0, \dots, 0)^{\top} \mbox{ and } D_{T^C}^*h=(0, \dots, 0,v_{k+1},\dots,v_d)^{\top}.
\end{equation}
We assume that the tight frame $D$ is normalized, i.e., $D D^* = I$ and $\|x\|_2=\|D^* x\|_2$ for all $x\in \R^p$.
Thus we have the following useful relation:
\begin{eqnarray}\label{eq:1.4}
\langle DD_T^* h, DD_{T^C}^*h\rangle &=& \langle DD_T^* h, DD^*h-DD_{T}^*h\rangle = \langle DD_T^*h, h\rangle-\|DD_{T}^*h\|_2^2\notag \\
&=& \langle D_T^*h, D^*h\rangle-\|DD_{T}^*h\|_2^2 = \|D_{T}^*h\|_2^2-\|DD_{T}^*h\|_2^2.
\end{eqnarray}

From the facts $\|D_{\Omega}^*h\|_1 \le \|D_{T}^*h\|_1$ and $\|D_{\Omega}^*h\|_1+\|D_{\Omega^C}^*h\|_1=  \|D_{T}^*h\|_1+\|D_{T^C}^*h\|_1=\|D^*h\|_1$,
the relation  $\|D^*_{\Omega^C}h\|_1\le 2\|D^*_{\Omega^C}\beta\|_1+\|D^*_{\Omega}h\|_1$ yields
\begin{equation}\label{eq:1.5}
\|D^*_{T^C}h\|_1\le 2\|D^*_{\Omega^C}\beta\|_1+\|D^*_{T}h\|_1
\end{equation}
Since $\|D^*_{T^C}h\|_{\infty} \le \frac{\|D^*_{T}h\|_1}k\le
      \frac{2\|D^*_{\Omega^C}\beta\|_1+\|D^*_{T}h\|_1}k$,
we can use lemma \ref{lem:1.1} to get the following $\ell_1$-invariant convex $k$-sparse decomposition of $D^*_{T^C}h$ which is the key ingredient of the proof:
\begin{equation}\label{eq:1.6}
D^*_{T^C}h = \sum_{t=1}^M x_t w_t,
\end{equation}
with each $w_t\in \R^{d}$ being $k$-sparse, $\|w_t\|_1=\|D^*_{T^C}h\|_1$ and
$\|w_t\|_{\infty}\le \frac{2\|D^*_{\Omega^C}\beta\|_1+\|D^*_{T}h\|_1}k$.
From this and the Cauchy-Schwartz inequality, we have immediately
\begin{equation}\label{eq:1.7}
 \sum_{t=1}^M x_t \|w_t\|_2 \le \frac{2\|D^*_{\Omega^C}\beta\|_1+\|D^*_{T}h\|_1}{\sqrt{k}}\le
 \frac{2\|D^*_{\Omega^C}\beta\|_1}{\sqrt{k}}+\|D^*_{T}h\|_2.
\end{equation}
By triangle inequality,  $\|D_{T^C}^*h\|_2\le  \sum_{t=1}^M x_t \|w_t\|_2$ holds and (\ref{eq:1.7}) implies
\begin{equation}\label{eq:1.8}
\|D_{T^C}^*h\|_2\le \frac{2\|D^*_{\Omega^C}\beta\|_1}{\sqrt{k}}+\|D^*_{T}h\|_2.
\end{equation}
Note that $\|\beta - \hat{\beta}\|_2^2=\|h\|_2^2=\|D^*h\|_2^2=\|D_{T^C}^*h\|_2^2+\|D_T^*h\|_2^2$ and $D^*\beta-(D^*\beta)_{\max (k)}=D^*_{\Omega^C}\beta$. In order
to prove the theorem, it suffices to show that there are constants $C_0', C_1'$ such that
\begin{equation}\label{eq:1.9}
\|D_T^*h\|_2 \le C_0'\varepsilon+C_1'\frac{\|D^*_{\Omega^C}\beta\|_1}{\sqrt{k}}.
\end{equation}
In fact, assuming (\ref{eq:1.9}) we get
\begin{eqnarray*}
\|h\|_2&=&\sqrt{\|D_T^*h\|_2^2+\|D_{T^C}^*h\|_2^2}\\
&\le &\sqrt{\bigg(C_0'\varepsilon+C_1'\frac{\|D^*_{\Omega^C}\beta\|_1}{\sqrt{k}}\bigg)^2+\bigg(\frac{2\|D^*_{\Omega^C}\beta\|_1}{\sqrt{k}}+\|D^*_{T}h\|_2\bigg)^2}\\
&\le & C_0'\varepsilon+C_1'\frac{\|D^*_{\Omega^C}\beta\|_1}{\sqrt{k}} +\frac{2\|D^*_{\Omega^C}\beta\|_1}{\sqrt{k}}+\|D^*_{T}h\|_2\\
&=& 2C_0'\varepsilon+2(C_1'+1)\frac{\|D^*_{\Omega^C}\beta\|_1}{\sqrt{k}}
\end{eqnarray*}

Now let us prove (\ref{eq:1.9}). Denote
\[
\Pi := \big| \langle \Phi DD^*_{T}h, \Phi h\rangle\big|=\big| \langle \Phi DD^*_{T}h,  \Phi DD^*h\rangle\big|.
\]
First, as $D_T^*h$ is $k$-sparse, hence $2k$-sparse. we have and $\delta_k\le \delta_{2k}$, we have
\begin{equation}\label{eq:1.10}
\Pi \le  \|\Phi DD^*_Th\|_2\|\Phi h\|_2 \le \sqrt{1+\delta_{2k}} \|D^*_Th\|_2 2  \varepsilon.
\end{equation}
On the other hand, as each $D_T^*h+w_t$ is $2k$-sparse, by following an approach similar to that in \cite{CZ1,XuXu} we have
\allowdisplaybreaks
\small{
\begin{eqnarray*}
\Pi &\stackrel{(\ref{eq:1.6})}{\ge}&  \langle \Phi DD^*_{T}h,  \Phi DD_T^*h +  \Phi DD_{T^C}^*h \rangle
= \sum_{t=1}^M x_t \langle \Phi DD^*_{T}h,  \Phi DD_T^*h +  \Phi D w_t \rangle\\
&=& \sum_{t=1}^M x_t \big\langle \big( \Phi DD^*_{T}h +\frac{1}2 \Phi Dw_t\big) -\frac{1}2 \Phi D w_t,
\big(\Phi DD^*_{T}h +\frac{1}2 \Phi D w_t\big) +\frac{1}2 \Phi D w_t\big\rangle\\
&=&\sum_{t=1}^M x_t \left(\big\| \Phi DD^*_{T}h +\frac{1}2 \Phi D w_t\big\|_2^2- \big\|\frac{1}2 \Phi D w_t\big\|_2^2\right)\\
&\ge & \sum_{t=1}^M x_t \left((1-\delta_{2k})\big\| DD^*_{T}h +\frac{1}2  Dw_t\big\|_2^2-(1+\delta_{2k})\big\|\frac{1}2 Dw_t\big\|_2^2\right)\\
&= & \big(1-\delta_{2k}\big)\|DD^*_{T}h\|_2^2 + (1-\delta_{2k}) \sum_{t=1}^M x_t\langle DD^*_{T}h, Dw_t \rangle
- \frac{1}2\delta_{2k}\sum_{t=1}^M x_t \|Dw_t\|_2^2 \\
&\stackrel{(\ref{eq:1.6})}{=}  & \big(1-\delta_{2k}\big)\|DD^*_{T}h\|_2^2 + (1-\delta_{2k}) \langle DD^*_{T}h, DD^*_{T^C}h \rangle
- \frac{1}2\delta_{2k}\sum_{t=1}^M x_t \|Dw_t\|_2^2 \\
&\stackrel{(\ref{eq:1.4})}{=} & \big(1- \delta_{2k}\big)\|DD^*_{T}h\|_2^2 + (1-\delta_{2k}) \left(\|D^*_{T}h\|_2^2 -\| DD^*_{T}h \|_2^2\right)
- \frac{1}2\delta_{2k}\sum_{t=1}^M x_t \|Dw_t\|_2^2 \\
&=& (1-\delta_{2k})\|D^*_{T}h\|_2^2- \frac{1}2\delta_{2k}\sum_{t=1}^M x_t \|Dw_t\|_2^2
\ge (1-\delta_{2k})\|D^*_{T}h\|_2^2 - \frac{1}2\delta_{2k}\sum_{t=1}^M x_t \|w_t\|_2^2 \\
&\ge & (1-\delta_{2k})\|D^*_{T}h\|_2^2- \frac{1}2\delta_{2k}\left(\frac{2\|D^*_{\Omega^C}\beta\|_1}{\sqrt{k}}+\|D^*_{T}h\|_2\right)^2\\
&= &  (1-\frac{3}2 \delta_{2k} )\|D^*_{T}h\|_2^2 -
\delta_{2k}\left(\frac{2\|D^*_{\Omega^C}\beta\|_1^2}{k}+\frac{2\|D^*_{\Omega^C}\beta\|_1\|D^*_{T}h\|_2}{\sqrt{k}}\right).
\end{eqnarray*}
}
Combining this with (\ref{eq:1.10}) we see that
\[
(1-\frac{3}2 \delta_{2k} )\|D^*_{T}h\|_2^2 -
\delta_{2k}\left(\frac{2\|D^*_{\Omega^C}\beta\|_1^2}{k}+\frac{2\|D^*_{\Omega^C}\beta\|_1\|D^*_{T}h\|_2}{\sqrt{k}}\right)\le
\sqrt{1+\delta_{2k}} \|D^*_Th\|_2 2\varepsilon.
\]
By making perfect square, we have
\scriptsize{
\[
\left(\|D^*_{T}h\|_2 -\bigg( \frac{2\sqrt{1+\delta_{2k}}}{3(\frac{2}3-\delta_{2k})}\varepsilon +\frac{2\delta_{2k}}{3(\frac{2}3-\delta_{2k})}\frac{\|D^*_{\Omega^C}\beta\|_1}{\sqrt{k}}\bigg)\right)^2
\le \bigg( \frac{2\sqrt{1+\delta_{2k}}}{3(\frac{2}3-\delta_{2k})}\varepsilon +\frac{2\delta_{2k}}{3(\frac{2}3-\delta_{2k})}\frac{\|D^*_{\Omega^C}\beta\|_1}{\sqrt{k}}\bigg)^2+
\bigg(\sqrt{\frac{2\delta_{2k}}{3(\frac{2}3-\delta_{2k})}}\frac{\|D^*_{\Omega^C}\beta\|_1}{\sqrt{k}}\bigg)^2,
\]
}\normalsize{}
which implies that
\scriptsize{
\[
\|D^*_{T}h\|_2 -\bigg( \frac{2\sqrt{1+\delta_{2k}}}{3(\frac{2}3-\delta_{2k})}\varepsilon +\frac{2\delta_{2k}}{3(\frac{2}3-\delta_{2k})}\frac{\|D^*_{\Omega^C}\beta\|_1}{\sqrt{k}}\bigg)
\le  \frac{2\sqrt{1+\delta_{2k}}}{3(\frac{2}3-\delta_{2k})}\varepsilon +\frac{2\delta_{2k}}{3(\frac{2}3-\delta_{2k})}\frac{\|D^*_{\Omega^C}\beta\|_1}{\sqrt{k}}+
\sqrt{\frac{2\delta_{2k}}{3(\frac{2}3-\delta_{2k})}}\frac{\|D^*_{\Omega^C}\beta\|_1}{\sqrt{k}}.
\]
}\normalsize{}
Finally we get (\ref{eq:1.9}):
\[
\|D^*_{T}h\|_2\le \frac{4\sqrt{1+\delta_{2k}}}{3(\frac{2}3-\delta_{2k})}\varepsilon+
\frac{4\delta_{2k}+\sqrt{6\delta_{2k}(\frac{2}3-\delta_{2k})}}{3(\frac{2}3-\delta_{2k})}\frac{\|D^*_{\Omega^C}\beta\|_1}{\sqrt{k}}.
\]

\end{proof}
\section*{Acknowledgement:} I would like to thank my academic supervisor, Professor Guangwu Xu, for his assistance in this proof. I
would also like to thank Bing Gao at the Chinese Academy of Sciences for pointing out an error in the earlier version of the note. The bound here is weaker than
that reported earlier.

\end{document}